\pdfoutput=1
\documentclass[10pt,a4paper]{article}
\usepackage[T2A]{fontenc}
\usepackage[utf8]{inputenc}
\usepackage[english]{babel}
\usepackage{amssymb,graphicx}
\usepackage{amsmath,amsfonts}
\usepackage{amsthm}
\usepackage{framed}
\usepackage{fullpage}
\usepackage[makeroom]{cancel}
\usepackage[export]{adjustbox} 
\usepackage{feynmf}
\usepackage{microtype}
\usepackage{hyperref}

\title{\textbf{Higgsed network calculus}}
\author{Yegor
  Zenkevich\thanks{yegor.zenkevich@gmail.com} \thanks{Current
    affiliations:\protect\\
    {\small\textit{SISSA, via Bonomea 265, 34136
         Trieste, Italy,}}\protect\\
     {\small\textit{INFN, Sezione di Trieste,}}\protect\\
    {\small\textit{IGAP, via Beirut 2/1, 34151 Trieste, Italy,}}\protect\\
    {\small\textit{ITEP, Bolshaya Cheremushkinskaya street 25, 117218
      Moscow, Russia,}}\protect\\
  {\small\textit{ITMP MSU, Leninskie gory 1, 119991 Moscow, Russia,}}\protect\\
  {\small\textit{MIPT, Institutskii pereulok 9, 141700, Dolgoprudny, Russia}}
   } 
  \\
  {\small\textit{Dipartimento di Fisica, Universit\`a di Milano
      Bicocca, Piazza della Scienza 3, I-20126 Milano, Italy,}}\\
  {\small\textit{INFN, sezione di Milano-Bicocca, I-20126 Milano, Italy,}}\\
  {\small\textit{ITEP, Moscow 117218, Russia}}} \date{}
\begin{document}
\maketitle
   \vspace{-45ex}
\begin{flushright}
  ITEP-TH-40/18
\end{flushright}
\vspace{35ex}

\begin{abstract}
  We introduce a formalism for describing holomorphic blocks of $3d$
  quiver gauge theories using networks of Ding-Iohara-Miki algebra
  intertwiners. Our approach is very direct and gives an explicit
  identification of the blocks with Dotsenko-Fateev type integrals for
  $q$-deformed quiver $W$-algebras. We also explain how quiver
  theories corresponding to Dynkin diagrams of superalgebras arise,
  write down the corresponding partition functions and $W$-algebras,
  and explain the connection with supersymmetric Macdonald-Ruijsenaars
  commuting Hamiltonians.
\end{abstract}

\section{Introduction}
\label{sec:introduction}
Ding-Iohara-Miki (DIM) algebra~\cite{DIM} is a unique and beautiful
object. It can be understood as a quantum affinization
$U_q(\widehat{\mathfrak{g}})$ of an algebra $\mathfrak{g}$ which is
\emph{itself} and affine algebra\footnote{The construction also works
  for $\mathfrak{g} = \widehat{\mathfrak{gl}}_n$.} $\mathfrak{g}=
\widehat{\mathfrak{gl}}_1$, deformed by an additional parameter
$t$. Because of the presence of \emph{two} loops in the construction,
the algebra is often called quantum \emph{toroidal,} and we will
denote it by $U_{q,t}(\widehat{\widehat{\mathfrak{gl}}}_1)$. DIM
algebra is symmetric under the exchange of \emph{three} parameters
$q$, $t^{-1}$ and $\frac{t}{q}$. It has two gradings $(d,d_{\perp})$,
two central charges $(\gamma, \gamma_{\perp})$ (again coming from two
loops in the construction) and also an interesting automorphism group
$SL(2,\mathbb{Z})$ which acts on them as doublets. We collect the
relevant definitions related to DIM algebra and its representations in
Appendix~\ref{sec:dim-algebra-its}.

In addition to being interesting from purely algebraic and
representation theoretic point of view, DIM algebra is extremely
relevant for physics. For example:
\begin{itemize}
\item It is the symmetry behind the AGT relation~\cite{Alday:2009aq}
  between instanton series of $4d$ $\mathcal{N}=2$ (and $5d$
  $\mathcal{N}=1$) gauge theories and $2d$
  CFTs~\cite{AFS},~\cite{AFS2},~\cite{Zenkevich:2014lca}. It is also
  the origin of the spectral duality, which exchanges the gauge groups
  at the nodes of a quiver gauge theory with the group corresponding
  to the Dynkin diagram of the
  quiver~\cite{Bao:2011rc},~\cite{Mironov:2012uh},~\cite{Mironov:2016cyq}.

\item It plays the central role in refined topological
  strings~\cite{Iqbal:2007ii}, where the central object of the
  formalism --- refined topological vertex --- can be identified with
  the intertwining operator of Fock representations of DIM
  algebra~\cite{AFS}. See Eq.~\eqref{eq:1} for an illustration of such
  a vertex/intertwiner. It also endows toric Calabi-Yau three-folds
  with an interesting integrable structure~\cite{Awata:2016mxc} and
  implies $(q,t)$-KZ difference equations for refined topological
  string amplitudes~\cite{Awata:2017cnz}.

\item DIM algebra provides a universal way to understand
  ``non-perturbative Ward identities'', or
  $qq$-characters~\cite{Nekrasov:2017gzb},~\cite{Kimura:2015rgi},~\cite{Mironov:2016yue},~\cite{Bourgine:2016vsq}
  for $4d$, $5d$ and $6d$ quiver gauge theories. Composing the
  trivalent intertwiners of Fock representations of DIM algebra
  according to a toric diagram of a CY threefold, one can build a
  ``two-dimensional'' (or network) matrix model, which is related to a
  family of Dotsenko-Fateev-like integral ensembles and the
  corresponding $q$-deformed vertex operator
  algebras~\cite{Morozov:2015xya}.
  
\item Very recently it was identified as the cohomological Hall
  algebra associated to CY three-folds~\cite{Gaiotto:2017euk}.

\item Higgsing construction can be employed to get holomorphic
  blocks~\cite{Pasquetti:2011fj} of $3d$ $\mathcal{N}=2^{*}$ gauge
  theories\footnote{By this we denote $\mathcal{N}=4$ gauge theories
    with supersymmetry softly broken by a real axial mass. For the
    details see~\cite{Zenkevich:2017ylb}.}  from a specifically tuned
  network of intertwiners of DIM algebra~\cite{Zenkevich:2017ylb}.
\end{itemize}

In the current paper we focus on the last item in the above list
(however, as we will see there are further ramifications) and
introduce a convenient formalism for describing $3d$
$\mathcal{N}=2^{*}$ quiver gauge theories.

Let us first recall the Higgsing construction (for details
see~\cite{Zenkevich:2017ylb}). The $3d$ theories can be understood as
worldvolume theories on the vortices in the Higgs phase of $5d$
$\mathcal{N}=1$ gauge theories~\cite{Hanany:2004ea}. In the
$\Omega$-background there is an analogue of geometric
transition~\cite{Dorey:2011pa}, which relates the theory with $M$
vortices in the Higgs phase to the theory \emph{without} vortices in
the Coulomb phase with scalar field vev $a = \epsilon_2 M$. The
geometric transition interpretation arises when we consider the Type
IIB brane construction of the $5d$ gauge theory. $M$ vortices on the
Higgs branch correspond to $M$ D3 branes stretching between NS5 and
D5' branes, while the dual side (after the transition) corresponds to
the resolution of the NS5 and D5' crossing. The five-brane picture, in
turn is related to the combination of DIM intertwiners, as shown
in~\cite{AFS}. This combination gives the partition function of the
Higgsed $5d$ theory, and thus of the $3d$ theory.

Our new formalism bypasses the complicated Higgsing procedure. The
major source of complications in the Higgsing approach is the need to
construct an \emph{auxiliary} $5d$ $\mathcal{N}=1$ gauge theory, then
tune its parameters to specific values, so that it reproduces the $3d$
theory on the worldvolume of the vortex defects appearing in the Higgs
branch of the $5d$ theory.

\begin{figure}[h]
  \centering
  \includegraphics{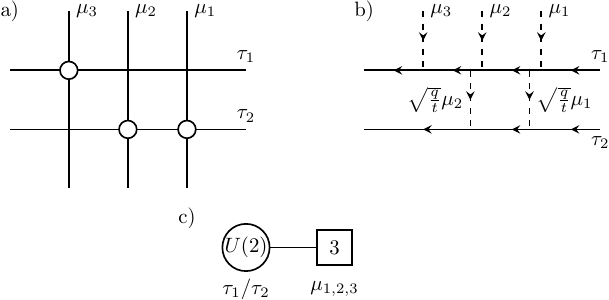}
  \caption{Comparison of two different formalisms for building a $3d$
    $\mathcal{N}=2^{*}$ quiver gauge theory. a) The old formalism:
    horizontal and vertical solid lines denote horizontal and vertical
    Fock representations respectively; circles and crossings denote
    resolved conifold-like geometries with specially tuned
    parameters. The notation is explained in detail
    in~\cite{Zenkevich:2017ylb}. b) The new formalism. Solid
    horizontal lines still denote horizontal Fock spaces, while
    vertical dashed lines are vertical \emph{vector}
    representations. They are joined together by Higgsed vertices.
    Notice how each circle in a) gives rise to dashed line emanating
    upwards from the corresponding point in b). Individual dashed
    lines correspond to screening charges acting on the Fock spaces
    represented by solid lines, so that the overall picture gives a
    Dotsenko-Fateev-like integral representation of the holomorphic
    block. c) The quiver gauge theory modelled by a) and b).}
  \label{fig:1}
\end{figure}

Our new formalism, which we call the \emph{Higgsed} network calculus,
avoids the intermediate step (the auxiliary $5d$ theory) and allows
for \emph{direct} computation of the holomorphic blocks of the $3d$
theories. The formalism employs a ``Higgsed'' vertex which resembles
refined topological vertex, but, unlike the latter, doesn't introduce
a bend in the five-brane. An example of two descriptions of the same
$3d$ theory using the old and the new formalism is shown in
Fig.~\ref{fig:1}.  Overall, the Higgsed network looks as a collection
of D3 branes (dashed lines in Fig.~\ref{fig:1}) stretched between a
stack of parallel five-branes (solid lines).

On the algebraic side, Higgsed vertices can be thought of as
elementary building blocks of \emph{screening currents,} which commute
with the action of a certain $W$-algebra and in this way can be used
to \emph{define} this algebra~\cite{Litvinov:2016mgi}~\cite{BFM}. It
also turns out that in our approach we can easily reproduce the
well-known result~\cite{Bullimore:2014awa} --- that the holomorphic
blocks of $3d$ gauge theories of the kind we are considering are
eigenfunctions (or, more generally, kernel functions) of trigonometric
Ruijsenaars-Schneider Hamiltonians.

Having thus reproduced the results of the old approach, we continue to
some generalizations which are natural in the new formalism. We
consider all \emph{three} possible species of horizontal Fock
representations and all three possible species of vertical vector
representations and heavily employ the $\mathfrak{S}_3$ symmetry of
the DIM algebra to build a network incorporating all of them (see
Appendix~\ref{sec:dim-algebra-its} for details on the representations
of DIM algebra). This corresponds to introducing \emph{several} sorts
of screening currents. The resulting model was considered
in~\cite{BFM}, and corresponds to a $W$-algebra associated with a
\emph{superalgebra.} We have thus obtained partition functions of $3d$
quiver theories corresponding to Dynkin diagrams of
superalgebras. These theories should be $3d$ uplifts of $2d$ theories
recently studied in~\cite{Nekrasov:2018gne} (see also earlier
work~\cite{Orlando:2010uu}). We also prove that their partition
functions are eigenfunctions of the \emph{supersymmetric}
Ruijsenaars-Schneider Hamiltonians~\cite{FS}.

The rest of the paper is structured as follows. In
sec.~\ref{sec:intertw-ruijs-hamilt} we present the systematic
introduction of our formalism and examples of the computations
reproducing the old construction: we write down one species of Higgsed
vertices, show how to compose them into screening currents and derive
the commutation relations for them. In sec.~\ref{sec:boson-ferm-scre}
we introduce the complete toolbox of vertices, build the
\emph{fermionic} screening current and write down the corresponding
partition functions. We prove that the network partition functions are
eigenfunctions of the supersymmetric Ruijsenaars-Schneider
Hamiltonians in sec.~\ref{sec:ruijs-hamilt}. We present our
conclusions and comment on future directions in
sec.~\ref{sec:concl-disc}.

\section{Intertwiners and blocks}
\label{sec:intertw-ruijs-hamilt}
In this section we introduce the Higgsed vertices $\Phi$ and
$\Phi^{*}$, from which we build the ``Higgsed network''. The vacuum
matrix element of the network will give the partition function
(holomorphic block) of the $3d$ quiver gauge theory.

We have collected the definitions of the DIM algebra and its relevant
representations in the Appendix~\ref{sec:dim-algebra-its} not to
clutter the presentation with too many technical details. However, the
reader who is not familiar with DIM formalism is invited to consult it
before proceeding to the main part of the text.

As a warm-up, let us recall the construction of the conventional
refined topological vertices as intertwining operators of DIM algebra,
proposed in~\cite{AFS}. In this approach five-branes of Type IIB
string theory (or, equivalently, the edges of the toric diagram of a
CY threefold) are identified with Fock representations
$\mathcal{F}^{(m,n),q,t^{-1}}_u$ of DIM algebra (see
sec.~\ref{sec:horiz-fock-repr} for the definition of a Fock
representation). The central charge vector of the Fock representation
corresponds to the type of the fivebrane with (depending on an
irrelevant choice of $SL(2,\mathbb{Z})$ duality frame) $(1,0)$ meaning
NS5 and $(0,1)$ meaning D5'. The intertwiner of Fock representation is
a trivalent junction of branes, as shown below:
\begin{equation}
  \label{eq:1}
  \includegraphics[valign=c]{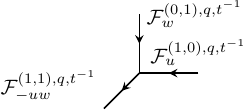}
\end{equation}
Since the charges of the representations, or branes are conserved, the
branes are bend at the junction. The basis in a Fock space is labelled
by Young diagrams. Therefore, solid lines in Eq.~\eqref{eq:1} each
carry a Young diagram, and gluing of two legs is performed by summing
over a complete basis of states in the Fock space, i.e.\ over all
Young diagrams. The result is just a network of intertwining
operators, composed according to a five-brane web (or toric diagram of
a CY). The vacuum matrix element of the network of intertwiners is
equal to the refined topological string partition function on the
toric CY~\cite{Mironov:2016yue} (see also ).

\subsection{The intertwiners $\Phi$ and $\Phi^{*}$}
\label{sec:intertwiners}
Let us introduce the main character of our story, the Higgsed vertex,
or the vector intertwiner $\Phi(w): \mathcal{F}^{(1,0),q,t^{-1}}_u
\otimes \mathcal{V}^q_w \to \mathcal{F}_{tu}^{(1,0),q,t^{-1}}$, which
we draw as
\begin{equation}
  \label{eq:2}
  \Phi(w) =\quad  \includegraphics[valign=c]{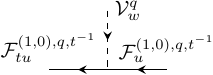}
\end{equation}

From now on all the Fock representations we will encounter will have
the ``direction'' of the central charge equal to $(1,0)$, and we omit
it from our notation. The dashed vertical lines denote the vertical
vector representations (see sec.~\ref{sec:vert-vect-repr}), while
solid horizontal lines are still horizontal Fock representations (see
sec.~\ref{sec:horiz-fock-repr}), the latter exactly the same as in the
ordinary DIM networks.

The vector representation has zero central charges, so when it joins a
Fock representation the five-brane is not bent.  This behavior reminds
one of a D3 brane ending on an NS5 brane. In a moment we will see that
this is not a coincidence and derive a precise relation between a
network of intertwiners and the partition function of the effective
$3d$ theory obtained from the D3 branes stretched between a stack of
five-branes.

The intertwining property of $\Phi$ means that $\Phi \Delta(g) = g
\Phi$ for any element $g$ of DIM algebra (see the definition of the
coproduct in sec.~\ref{sec:coproduct}). One can check that the
following explicit expression\footnote{We rescale the coordinate $w$
  by $q^{-\frac{1}{2}}$ compared to~\cite{CP1}.}~\cite{CP1} for the
intertwiner~(\ref{eq:2}) indeed satisfies this constraint:
\begin{equation}
  \label{eq:7}
  \Phi (w) = e^{- \epsilon_2 Q} w^{\frac{P}{\epsilon_1}} \exp \left[ - \sum_{n\geq 1} \frac{w^n}{n}
    \frac{1-t^{-n}}{1-q^n} a_{-n}  \right] \exp \left[ \sum_{n\geq 1} \frac{w^{-n}}{n}
    \frac{1-t^n}{1-q^{-n}} a_n \right]
\end{equation}
where we define $q = e^{\epsilon_1}$, $t = e^{\beta \epsilon_1}=
e^{-\epsilon_2}$ and the definition of the Fock space operators $a_n$,
$P$ and $Q$ are given in Eqs.~\eqref{eq:8},~\eqref{eq:13} in the
Appendix~\ref{sec:dim-algebra-its}. Notice that though the Fock space
is not bent (its slope is horizontal all the way through the
intertwiner), as we see from Eq.~\eqref{eq:2} and the explicit
expression~\eqref{eq:7}, its spectral parameter (which plays the role
of the position of the five-brane) is \emph{shifted} after it passes
the junction with the incoming dashed line.

We are already able to study the simplest nontrivial example, a
network of two intertwiners:
\begin{equation}
  \label{eq:9}
  \includegraphics[valign=c]{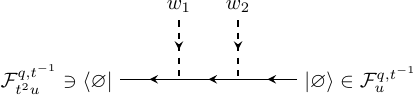} 
\end{equation}
The corresponding expression is the matrix element of the product of
two operators~(\ref{eq:7}), which is evaluated by normal ordering the
free fields:
\begin{multline}
  \label{eq:11}
  (\ref{eq:9}) = \langle t^2u, \varnothing | \Phi(w_1) \Phi(w_2) | u,
  \varnothing \rangle = w_1^{\log_q u + \beta} w_2^{\log_q u}
  \exp\left[ -\sum_{n\geq 1} \frac{1}{n} \left( \frac{w_2}{w_1}
    \right)^n \frac{1 -
      t^{-n}}{1-q^{-n}}\right] =\\
  = w_1^{\log_q u + \beta} w_2^{\log_q u} \frac{\left( \frac{q}{t}
      \frac{w_2}{w_1} ;q \right)_{\infty}}{\left( q \frac{w_2}{w_1} ;q
    \right)_{\infty}}.
\end{multline}
Moving one step further, we get an answer for the network of $n$
operators $\Phi$ glued together horizontally:
\begin{equation}
  \label{eq:10}
  \includegraphics[valign=c]{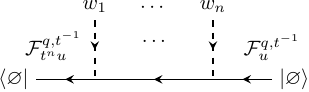}=
  A(\vec{w}) \left(
    \prod_{i=1}^n w_i^{\log_q u + \beta(i-1)} \right)  \prod_{i<j}
  \frac{\left( 
      \frac{w_i}{w_j} ;q \right)_{\infty}}{\left( t \frac{w_i}{w_j} ;q
    \right)_{\infty}}
\end{equation}
where we extract the $q$-periodic prefactor
\begin{equation}
  \label{eq:37}
  A(\vec{w})=\prod_{i<j}
  \left[ \left( \frac{w_i}{w_j} \right)^{\beta}  
  \frac{\theta_q\left( t
      \frac{w_i}{w_j}  \right)}{\theta_q\left( \frac{w_i}{w_j} 
    \right)} \right]
\end{equation}
from the matrix element; here $\theta_q(x) =
(q;q)_{\infty}(x;q)_{\infty}\left(\frac{q}{x};q\right)_{\infty}$ is
Jacobi theta-function. We will see in what follows that such
$q$-periodic factors will be mostly immaterial to the structure of the
network. In particular, they \emph{factor out} of the sums when we
glue vector states together, and give an overall $q$-periodic
prefactor for the network.

To build \emph{networks} of intertwiners similar to (but simpler than)
those considered
in~\cite{AFS},~\cite{Mironov:2016yue},~\cite{Bourgine:2016vsq} we need
one more operator, the dual intertwiner $\Phi^{*}(y):
\mathcal{F}_u^{(1,0)} \to \mathcal{F}_{u/t}^{(1,0)} \otimes
\mathcal{V}^q_y$, which satisfies $\Delta(g) \Phi^{*} = \Phi^{*} g
$. An explicit check shows that the operator
\begin{multline}
  \label{eq:5}
  \includegraphics[valign=c]{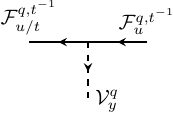}\quad =
  \Phi^{*}(y) =\\
  = e^{\epsilon_2 Q} y^{\beta - \frac{P}{\epsilon_1}} \exp \left[ \sum_{n\geq 1} \frac{y^n}{n}
    \left( \frac{t}{q} \right)^{\frac{n}{2}} \frac{1-t^{-n}}{1-q^n}
    a_{-n} \right] \exp \left[ - \sum_{n\geq 1} \frac{y^{-n}}{n}
    \left( \frac{t}{q} \right)^{\frac{n}{2}} \frac{1-t^n}{1-q^{-n}}
    a_n \right]
\end{multline}
satisfies the intertwining property. Notice how the spectral parameter
of the Fock space to the left of the dual intertwiner $\Phi^{*}$ is
shifted the opposite way, compared to the intertwiner $\Phi$.

The matrix elements of products of $\Phi^{*}$ are given by
\begin{equation}
  \label{eq:12}
\includegraphics[valign=c]{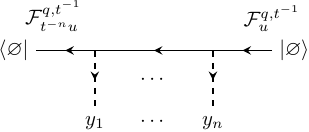}\quad =  \left(
  \prod_{i=1}^n y_i^{-\log_q u + \beta (n-i+1)}
\right) \prod_{i<j}
  \frac{\left(
    \frac{y_j}{y_i} ;q \right)_{\infty}}{\left( t \frac{y_j}{y_i} ;q
  \right)_{\infty}}.
\end{equation}
Combining $\Phi$ and $\Phi^{*}$ we get
\begin{multline}
  \label{eq:14}
  \includegraphics[valign=c]{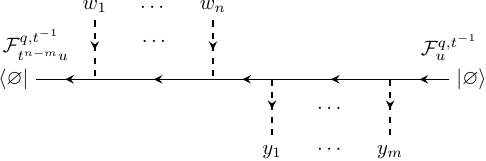} \quad = A(\vec{w}) B(\vec{w},\vec{y}) \left(
    \prod_{i=1}^n w_i^{\log_q u + \beta(i-2m-1)} \right)  \times\\
\times  \left(
  \prod_{i=1}^m y_i^{-\log_q u + \beta (i-2n+1)}
\right) \prod_{k<l}^n \frac{\left( \frac{w_k}{w_l} ;q
    \right)_{\infty}}{\left( t \frac{w_k}{w_l} ;q \right)_{\infty}}
  \prod_{i<j}^m\frac{\left( \frac{y_j}{y_i} ;q
    \right)_{\infty}}{\left( t \frac{y_j}{y_i} ;q \right)_{\infty}}
  \prod_{a=1}^m \prod_{b=1}^n \frac{\left( t \sqrt{\frac{q}{t}}
      \frac{w_b}{y_a} ;q \right)_{\infty}}{\left( \sqrt{\frac{q}{t}}
      \frac{w_b}{y_a} ;q \right)_{\infty}},
\end{multline}
where an additional $q$-periodic prefactor reads
\begin{equation}
  \label{eq:36}
  B(\vec{w},\vec{y}) =  \prod_{a=1}^m \prod_{b=1}^n \left[\left(
    \frac{y_a}{w_b} \right)^{\beta} \frac{\theta_q\left( t \sqrt{\frac{q}{t}}
      \frac{y_a}{w_b} \right)}{\theta_q\left( \sqrt{\frac{q}{t}}
      \frac{y_a}{w_b} \right)}\right].
\end{equation}

We can already notice that the $q$-Pochhammer factors in
Eq.~\eqref{eq:14} resemble those of the holomorphic block integrand
for a pair of $\mathcal{N}=2$ bifundamental chiral multiplets charged
under $U(n)\times U(m)$ flavour symmetry. In this case the parameter
$q$ is identified with the parameter of the $3d$ $\Omega$-background
$S^2 \times_q D_2$ and $t$ is related to the real axial mass
deformation of the $\mathcal{N}=4$ theory. Together the two chirals
constitute what we might call a bifundamental $\mathcal{N}=2^{*}$
multiplet, i.e.\ what remains of the $\mathcal{N}=4$ bifundamental
multiplet after turning on the $t$-parameters responsible for the soft
breaking of supersymmetry. Indeed, if we set the masses associated to
$U(n)$ flavours to $\mu_i = y_i$, $i=1,\ldots, n$ and the $U(m)$
masses to $\bar{\mu}_j = \sqrt{\frac{q}{t}} w_j$, $j=1,\ldots,m$, we
find that we reproduce the bifundamental contribution
\begin{equation}
  \label{eq:38}
  \prod_{a=1}^m \prod_{b=1}^n \frac{\left( t
      \frac{\bar{\mu}_b}{\mu_a} ;q \right)_{\infty}}{\left( 
      \frac{\bar{\mu}_b}{\mu_a} ;q \right)_{\infty}}.
\end{equation}
to the $D_2 \times_q S^1$ partition function. Thus, $n$ dashed lines
coming from the top of the picture in Eq.~\eqref{eq:14} correspond to
a $U(n)$ flavour group and $m$ dashed lines escaping from the bottom
are related to $U(m)$ flavour group. Thus, our initial guess that the
dashed lines are somehow associated with D3 branes seem to be
plausible: the $3d$ bifundamental chirals couple the $U(n)$ and $U(m)$
gauge theories living on the stacks of D3 branes. The D3 branes are
semi-infinite, so the gauge fields are frozen and the gauge symmetries
become flavour symmetries of the $3d$ theory of bifundamental chirals.

What is the field theory role of the remaining factors in
Eq.~\eqref{eq:14}? First of all, the $q$-periodic contributions in the
holomorphic blocks are not important, since they cancel when one
combines two blocks into a partition function for a compact manifold
(e.g.\ $S^3_b$). Additional $q$-Pochhammers in Eq.~\eqref{eq:14} can
be thought of as coming from \emph{flipping
  fields}~\cite{Zenkevich:2017ylb}, charged under flavour symmetries
$U(n)$ and $U(m)$.

We will see in what follows that when we \emph{glue} pictures
like~\eqref{eq:14} along the dashed lines we effectively \emph{gauge}
the corresponding flavour groups. The flipping fields coming from both
sides then combine into an $\mathcal{N}=2$ vector and adjoint chiral
contribution for the gauged symmetry. This is in accordance with the
D3 brane interpretation: the couplings on the branes are unfrozen,
when the branes have \emph{finite} length.

\subsection{Commutation relations for the intertwiners}
\label{sec:comm-relat-intertw}
For the moment we have only shown how to compose the intertwiners
$\Phi$ and $\Phi^{*}$ horizontally. However, we can already make a
natural and meaningful exercise with our building blocks. Let us
compare different orderings of the intertwiners along the solid
line. There are three possibilities:
\begin{enumerate}
\item Commutation of $\Phi$ with $\Phi$:
  \begin{equation}
  \label{eq:15}
  \includegraphics[valign=c]{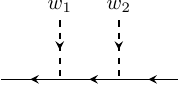} \quad = \left[
    \left(\frac{w_1}{w_2} \right)^{\beta} \frac{\theta_q \left( t \frac{w_1}{w_2} \right)}{\theta_q \left(
    \frac{w_1}{w_2} \right)} \right] R \left( \frac{w_1}{w_2} \right) \times \quad \includegraphics[valign=c]{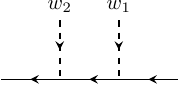}
\end{equation}
where
\begin{equation}
  \label{eq:16}
  R \left( x \right) =  \frac{\left(
      x ;q \right)_{\infty} \left( q
      x ;q \right)_{\infty}}{\left( t x ;q
    \right)_{\infty} \left( \frac{q}{t} x ;q
    \right)_{\infty}}.
\end{equation}
The terms in the square brackets in Eq.~\eqref{eq:15} combine into a
$q$-periodic function of $w_{1,2}$, which, as we have mentioned above,
is not important for our network construction. The function $R(x)$ is
the ``miniature version'' of the DIM
$R$-matrix~\cite{Awata:2016mxc},~\cite{FJMM-R}. In our case the
$R$-matrix permutes two vector representations living on the vertical
dashed lines.

\item Commutation of $\Phi^{*}$ with $\Phi^{*}$. For the dual
  intertwiners we find
\begin{equation}
  \label{eq:17}
  \includegraphics[valign=c]{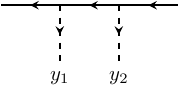} \quad =
  \left[\left( \frac{y_1}{y_2} \right)^{\beta}  \frac{\theta_q \left( \frac{y_2}{y_1} \right)}{\theta_q \left(
      t \frac{y_2}{y_1} \right)} \right] \frac{1}{R \left( \frac{y_1}{y_2} \right)} \times \quad \includegraphics[valign=c]{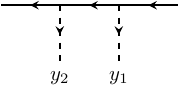}
\end{equation}
which features an inverse of the $R$-matrix from Eq.~\eqref{eq:16}
together with another $q$-periodic factor.

\item Commutation of $\Phi$ with $\Phi^{*}$. Finally, we have
  \begin{equation}
  \label{eq:18}
  \includegraphics[valign=c]{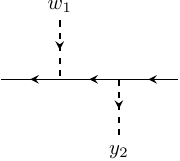} \quad =
  \left[\left( \frac{y_2}{w_1} \right)^{\beta}  \frac{\theta_q \left(
        \sqrt{\frac{q}{t}} t \frac{y_2}{w_1} \right)}{\theta_q \left(
      \sqrt{\frac{q}{t}} \frac{y_2}{w_1} \right)} \right] \times \quad \includegraphics[valign=c]{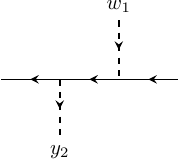}
\end{equation}
We find that $\Phi$ and $\Phi^{*}$ \emph{commute,} up to $q$-periodic
factors, which for us is as good as commutativity.
\end{enumerate}

\subsection{Vertical gluing and $q$-Virasoro screening charges}
\label{sec:trace-over-vector}
We pass to the next necessary step in building the network of
intertwiners --- vertical gluing. In our convention the
states in the vector representation correspond to the shifts of the
spectral parameter $w \mapsto q^k w$. Thus, to sum over the complete
basis of states in the vector representation we need to take the sum
over $k$, or, equivalently, the so-called Jackson $q$-integral over
the spectral parameter $w$ of the vector representation. For example,
gluing together two dual intertwiners we get an operator
$\mathcal{Q}^{q,t^{_1}}_q: \mathcal{F}^{q,t^{-1}}_{u_1} \otimes
\mathcal{F}^{q,t^{-1}}_{u_2} \to \mathcal{F}^{q,t^{-1}}_{t^{-1} u_1}
\otimes \mathcal{F}^{q,t^{-1}}_{t u_2}$:
\begin{multline}
  \label{eq:20}
  \mathcal{Q}_q^{q,t^{-1}} = \includegraphics[valign=c]{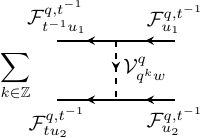}
  \quad = \sum_{k \in \mathbb{Z}}
  \begin{array}{c}
    \Phi^{*}(q^kw)\\
    \otimes\\
    \Phi(q^kw)
  \end{array} = \int_{-\infty}^{\infty} d_q w
  \begin{array}{c}
    \Phi^{*}(w)\\
    \otimes\\
    \Phi(w)
  \end{array} = \int_{-\infty}^{\infty} d_q w\, S_q^{q, t^{-1}}(w) = \\
  = \sum_{k \in \mathbb{Z}}
  e^{-\epsilon_2 (Q_1-Q_2)} \left( q^k w \right)^{\beta + \frac{P_1-P_2}{\epsilon_1}} \exp \left[ -
    \sum_{n\geq 1} \frac{w^n}{n} q^{nk}
    \frac{1-t^{-n}}{1-q^n} \left( a^{(2)}_{-n} -
      \left(\frac{t}{q}\right)^{\frac{n}{2}} a^{(1)}_{-n}  \right)
  \right]\times\\
  \times \exp \left[ \sum_{n\geq 1} \frac{w^{-n}}{n}
q^{-nk}    \frac{1-t^n}{1-q^{-n}} \left( a_n^{(2)} -
      \left(\frac{t}{q}\right)^{\frac{n}{2}} a^{(1)}_{n} \right) \right],
\end{multline}
where we denote by $a^{(1)}_n$ (resp.\ $a^{(2)}_n$) creation and
annihilation operators acting on the upper (resp.\ lower) horizontal
Fock space (similarly for the zero modes $P_{1,2}$ and $Q_{1,2}$). One
more convenient representation of the vertical gluing is the
\emph{contour} integral over an appropriate contour in the complex $w$
plane. One way to rewrite Eq.~\eqref{eq:20} as a contour integral is
to notice that
\begin{equation}
  \label{eq:29}
  \frac{(q;q)_{\infty}^2 }{(a;q)_{\infty}\left(\frac{q}{a};q\right)_{\infty}} \oint_{\mathcal{C}} \frac{d\xi}{\xi} \left(\frac{\xi}{w}\right)^{\log_q a}
  \frac{\theta_q \left(\frac{a\xi}{w} \right)}{\theta_q \left(\frac{\xi}{w}\right)} f(\xi) =  \sum_{k \in \mathbb{Z}}
  f(q^k w),
\end{equation}
for almost any $a \in \mathbb{C}$, where $\mathcal{C}$ wraps the poles
of the theta-function in the denominator. Therefore, by inserting the
theta-functions and a prefactor, as in Eq.~\eqref{eq:29}, under the
integral and integrating over a specific contour we can turn the
integral into a sum. It turns out that for certain combinations of
intertwiners the contour $\mathcal{C}$ can be traded for other
contours wrapping poles of the correlation functions~\eqref{eq:14}.
As have been noted in e.g.\ \cite{Zenkevich:2014lca, Mironov:2016cyq},
eventually all the representations of the ``trigonometric integrals''
--- either as a sum, or as a Jackson $q$-integral, or as a contour
integral --- boil down to the same expression and are completely
equivalent. Similar equivalence occurs in the Dotsenko-Fateev
representations of $q$-deformed conformal blocks and in the
holomorphic blocks of $3d$ theories. We will mostly use the contour
integral representation, usually assuming that the contour wraps the
poles of the correlation functions under the integral.

In fact, the \emph{integrand} $S_q^{q, t^{-1}}(w)$ in the first line
of Eq.~\eqref{eq:20} is nothing but the \emph{screening current} of
the $q$-Virasoro algebra $\mathsf{Vir}_{q,t}$, built from the pair of
free bosons $a^{(1,2)}_n$~\cite{Mironov:2016yue}. Thus
$\mathcal{Q}^{q,t^{-1}}_q$ is the screening charge, commuting with the
action of $\mathsf{Vir}_{q,t}$.  As explained
in~\cite{FJMM-plane},~\cite{Mironov:2016yue} this algebra is a
\emph{subalgebra} of the DIM algebra, i.e.\ one can build a current
from a combination of DIM generators, so that when acting on the
tensor product of two horizontal Fock representations it reproduces
the relations of $\mathsf{Vir}_{q,t}$. Since the network of
intertwiners~\eqref{eq:20} commutes with $\Delta(g)$ for any $g$ from
the DIM algebra, it also commutes with the $q$-Virasoro current. We
can write the intertwining relation graphically:
\begin{equation}
  \label{eq:30}
  \includegraphics[valign=c]{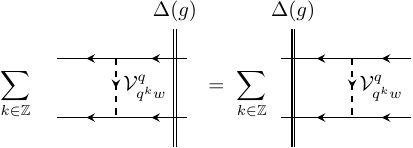}
\end{equation}
where the double line denotes the position of the operator $\Delta(g)$
on the horizontal lines, or as a formula
\begin{equation}
  \label{eq:31}
  [ \mathcal{Q}_q^{q,t^{-1}} , \Delta(g)] = 0.
\end{equation}
A more general network can be obtained by adding external dashed
lines, e.g.\
\begin{equation}
  \label{eq:32}
  \includegraphics[valign=c]{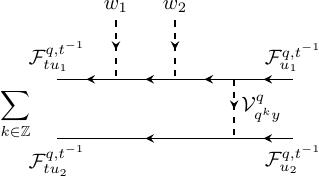}
\end{equation}
One can see that the external lines correspond to \emph{degenerate
  vertex operators} of the $\mathsf{Vir}_{q,t}$ algebra, similarly
to~\cite{Mironov:2016yue}. The
network~\eqref{eq:32} is therefore a product of \emph{screened}
degenerate vertex operators. Taking the vacuum matrix element of
Eq.~\eqref{eq:32} we get
\begin{multline}
  \label{eq:33}
  \begin{array}{c}
    \langle \varnothing |\\
    \otimes\\
    \langle \varnothing |
  \end{array} \Bigl( \ref{eq:32} \Bigr) \begin{array}{c}
    |\varnothing \rangle\\
    \otimes\\
    |\varnothing \rangle
  \end{array} =  A(w_1, w_2) B(w_1, w_2, y) \left(
    \prod_{i=1}^2 w_i^{\log_q u + \beta(i-3)} \right) \times\\
  \times  \frac{\left( \frac{w_1}{w_2} ;q
    \right)_{\infty}}{\left( t \frac{w_1}{w_2} ;q
    \right)_{\infty}} \int_{-\infty}^{\infty} d_qy \,    y^{\log_q \frac{u_2}{u_1} -2 \beta}
   \prod_{i=1}^2 \frac{\left( t
      \sqrt{\frac{q}{t}} \frac{w_i}{y} ;q \right)_{\infty}}{\left(
      \sqrt{\frac{q}{t}} \frac{w_i}{y} ;q \right)_{\infty}}.
\end{multline}
where $A(\vec{w})$ and $B(\vec{w},\vec{y})$ are periodic factors from
Eq.~\eqref{eq:37} and from Eq.~\eqref{eq:36} respectively.

The $3d$ gauge theory corresponding to the network~\eqref{eq:32} can
be deduced either from the form of the integrand~\eqref{eq:33}
($q$-Pochhammer ratios give two pairs of chirals and a single
integration implies a $U(1)$ gauge group) or directly by
interpreting~\eqref{eq:32} as a brane picture (the intermediate dashed
line gives the $U(1)$ gauge theory and two upper dashed lines
contribute two fundamental multiplets, i.e.\ two pairs of fundamental
chirals). In any case, the resulting theory is the $\mathcal{N}=2^{*}$
version of the $T[SU(2)]$ theory. The masses of the fundamental
multiplets are $\mu_i = \sqrt{\frac{q}{t}} w_i$ and the FI parameter
of the $U(1)$ gauge group is $\tau = \frac{u_2}{u_1}$. The prefactor
in the second line of Eq.~\eqref{eq:33} corresponds to flipping fields
of the $U(2)$ flavour symmetry group.

Let us make a comment about the contours, over which one can integrate
the screening currents. If we consider the space of intertwining
operators of the form~\eqref{eq:32} as a vector space over
$q$-periodic functions of the parameters, it turns out to be two
dimensional. One can obtain this fact using several different lines of
arguments:
\begin{enumerate}
\item As we show in sec.~\ref{sec:ruijs-hamilt} the
  network~\eqref{eq:32} is an eigenfunctions of the Ruijsenaars
  difference operator with the eigenvalue independent of
  $y$. $q$-periodic functions pass through the difference operator, so
  one can study the space of eigenfunctions as a vector space over the
  field of $q$-periodic functions in the same way as one studies the
  space of eigenfunctions of a differential operator over the field of
  constants. The difference operator is of second order, therefore the
  space of solutions is two-dimensional.

\item The $3d$ theory $T[SU(2)]$ corresponding to the network has two
  vacua. These vacua correspond to two linearly independent networks
  of intertwiners.
\end{enumerate}
We can write two linearly independent networks of intertwiners
similarly to~\eqref{eq:33}, but with the Jackson $q$-integrals
replaced by the contour integrals over the contours
$\mathcal{C}_{1,0}$ and $\mathcal{C}_{0,1}$, wrapping the poles at $y
= \sqrt{\frac{q}{t}} w_1 q^k$, $k \in \mathbb{Z}_{\geq 0}$ and $y =
\sqrt{\frac{q}{t}} w_2 q^k$, $k \in \mathbb{Z}_{\geq 0}$
respectively. One can also isolate the contributions of contours
$\mathcal{C}_{1,0}$ and $\mathcal{C}_{0,1}$ by taking the residue in
the $y$-parameter of the intermediate dashed line at $y =
\sqrt{\frac{q}{t}} w_{1,2}$.

In what follows, not to overburden the notation, we will omit writing
the sums and the shifts of the intermediate dashed lines on our
diagrams. In this convention two networks corresponding to contours
$\mathcal{C}_{1,0}$ and $\mathcal{C}_{0,1}$ are given by the following
pictures
\begin{equation}
  \label{eq:35}
  \includegraphics[valign=c]{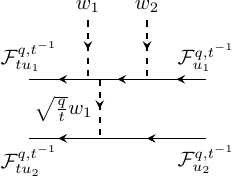}\qquad \qquad \qquad  \includegraphics[valign=c]{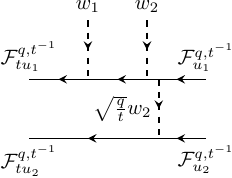}
\end{equation}
Notice how the lower dashed lines ``cling'' to the upper dashed
lines. As we have explained, this is the consequence of the structure
of poles of the integrand. For several screenings stretched between
two neighbouring horizontal lines (i.e.\ for a non-abelian $3d$
theory), from the first look at the structure of the integrand one
could have assumed that the intermediate lines can also ``cling''
together forming stacks of branes. Indeed, the interaction between two
parallel branes gives a denominator $\left( t \frac{y_k}{y_l} ;q
\right)_{\infty}$. For example, the picture:
\begin{equation}
  \label{eq:39}
  \includegraphics[valign=c]{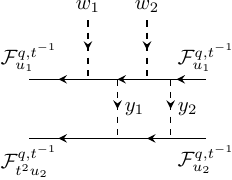}
\end{equation}
corresponds to the integral
\begin{equation}
  \label{eq:40}
  \oint_{\mathcal{C}} \frac{d^2y}{y_1 y_2}    (y_1 y_2)^{\log_q
    \frac{u_2}{u_1} -2 \beta}    \prod_{k\neq l}^2 \frac{\left(  \frac{y_k}{y_l} ;q \right)_{\infty}}{\left(
      t \frac{y_k}{y_l} ;q \right)_{\infty}}
  \prod_{i=1}^2 \prod_{j=1}^2 \frac{\left( t
      \sqrt{\frac{q}{t}} \frac{w_i}{y_j} ;q \right)_{\infty}}{\left(
      \sqrt{\frac{q}{t}} \frac{w_i}{y_j} ;q \right)_{\infty}}.
\end{equation}
The this ``stacking'' of branes would have corresponded to the contour
of double integration $\mathcal{C} = \mathcal{C}_{0,2}$ over the
spectral parameters $y_1$ and $y_2$, which wraps the poles $y_1 =
\sqrt{\frac{q}{t}} w_2 q^{k_1}$, $y_2 = \sqrt{\frac{q}{t}} w_2 q^{k_2}
t^{-1}$, with $k_{1,2} \in \mathbb{Z}_{\geq 0}$. However, in fact this
group of poles gets cancelled by the numerators $\left( t
  \sqrt{\frac{q}{t}} \frac{w_i}{y_j} ;q \right)_{\infty}$ in the
integrand~\eqref{eq:40}. Therefore parallel dashed lines between the
same horizontal lines \emph{cannot} cling together and always cling to
\emph{separate} upper dashed lines.

As one can deduce from the brane diagram, the $3d$ theory
corresponding to the network~\eqref{eq:39} is the $U(2)$ gauge theory
(two intermediate dashed lines) with two fundamental multiplets (two
external dashed lines). It has only one arrangement of branes in which
$y_{1,2} = \sqrt{\frac{q}{t}} w_{1,2} q^{k_{1,2}}$. Thus, the
resulting theory has only one vacuum. Geometrically, this theory
corresponds to counting certain quasimaps to the target
$\mathrm{Gr}(2,2) \simeq \mathrm{Gr}(0,2)$, i.e.\ to a point.

In general, the network with two horizontal lines corresponds to a
$3d$ with a single gauge group $U(m)$ and a number $n$ of fundamental
multiplets:
\begin{equation}
  \label{eq:41}
  \includegraphics[valign=c]{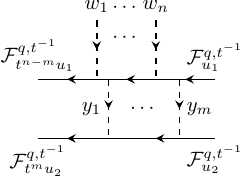}\qquad \Leftrightarrow \qquad   \includegraphics[valign=c]{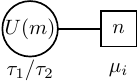}
\end{equation}
There are $\frac{n!}{m!(n-m)!}$ vacua, corresponding to the contours
of integration $\mathcal{C}_{1,0,\ldots, 1}$ with $m$ ones and $(n-m)$
zeroes. Graphically this number is explained as follows: $m$
intermediate dashed lines can cling to separate external dashed line,
and all the intermediate dashed lines are identical, hence $m!$ in the
denominator. These vacua are identified with fixed points of
$(\mathbb{C}^{*})^n$ action on $\mathrm{Gr}(m,n)$ (notice that we
consider only $m \leq n$). The corresponding integrals are given by
\begin{equation}
  \label{eq:42}
 \begin{array}{c}
    \langle \varnothing |\\
    \otimes\\
    \langle \varnothing |
  \end{array} \Bigl(  \ref{eq:41} \Bigr) \begin{array}{c}
    |\varnothing \rangle\\
    \otimes\\
    |\varnothing \rangle
  \end{array} \sim \prod_{k<l}^n \frac{\left( \frac{w_k}{w_l} ;q
    \right)_{\infty}}{\left( t \frac{w_k}{w_l} ;q
    \right)_{\infty}} \oint_{\mathcal{C}_{k_1,\ldots, k_n}} d^my \prod_{i=1}^m y^{\log_q
    \frac{u_2}{u_1} - 2 \beta - 1} \frac{\Delta_m^{(q,t)}(\vec{y})}{\bar{\Delta}_{m,n}^{(q,t)}(\vec{y},\vec{w})},
\end{equation}
where
\begin{equation}
  \label{eq:43}
  \Delta^{(q,t)}_m(\vec{y}) = \prod_{k\neq l}^m \frac{\left(  \frac{y_k}{y_l} ;q \right)_{\infty}}{\left(
      t \frac{y_k}{y_l} ;q \right)_{\infty}},
  \qquad \bar{\Delta}^{(q,t)}_{m,n}(\vec{y},\vec{w}) =  \prod_{i=1}^n \prod_{j=1}^m \frac{\left(
      \sqrt{\frac{q}{t}} \frac{w_i}{y_j} ;q \right)_{\infty}}{\left(
      t \sqrt{\frac{q}{t}} \frac{w_i}{y_j} ;q \right)_{\infty}}.
\end{equation}
This is precisely the integral for the holomorphic block of the $U(m)$
theory with $n$ fundamental multiplets. The prefactors again play the
role of flipping fields of the flavour symmetry. We can relate the
parameters of the network to that of the gauge theory on $S^1 \times_q
\mathbb{R}^2$. We write down the dictionary in Tab.~\ref{tab:1}.
\begin{table}[h]
  \centering
  \begin{tabular}[h]{|c|c|}
    \hline
    $3d$ $U(m)$ gauge theory & Higgsed network~\eqref{eq:32}\\
    \hline
    Axial real mass $m_A$ & $\frac{1}{R_{S^1}}\ln \left(-\frac{\sqrt{q}}{t}\right) $\\
    $3d$ $\Omega$-background parameter & $q$\\
    FI parameter $\tau$ & $\frac{u_2}{u_1}$\\
    Flavour masses $\mu_i$ & $\sqrt{\frac{q}{t}} w_i$\\
    Vacua & Contours of integration\\
    \hline
\end{tabular}
\caption{Dictionary between the  parameters of the $3d$ gauge theory and Higgsed
  network.}
  \label{tab:1}
\end{table}

\subsection{$qW_N$-algebra screenings and $3d$ quivers}
\label{sec:qw-algebra-scre}
We can stack more than two horizontal Fock representations on top of
each other and stretch vector representations between them. For example:
\begin{equation}
  \label{eq:21}
   \includegraphics[valign=c]{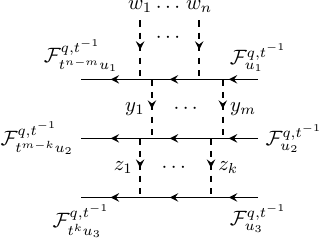}
\end{equation}
The intermediate dashed lines can now stretch between either the upper
two Fock spaces or the lower two. This gives rise to \emph{two}
screening currents, $S^{q,t^{-1}}_{q,12}(y)$ and
$S^{q,t^{-1}}_{q,23}(z)$ (and the corresponding screening charges),
which are similar to the Virasoro one \eqref{eq:20} with appropriate
change of $a^{(1,2)}_n$ to $a^{(2,3)}_n$ in
$S^{q,t^{-1}}_{q,23}(z)$. The algebra commuting with the screening
charges is the $q$-deformed $W_3$-algebra. The commutation can be
inferred from the DIM intertwining relations exactly as in
Eq.~\eqref{eq:30}.

Two sorts of screening currents have nontrivial normal ordering,
because the bosonic operators $a^{(2)}_n$ are shared between them. In
fact the normal ordering produces the interaction, dictated by the
$A_2$ Cartan matrix, between the integration variables $y_i$ and
$z_j$. Thus, the vacuum matrix element of the network~\eqref{eq:21} is
the $A_2$-type $q$-conformal matrix model, as
in~\cite{Mironov:2016cyq}:
\begin{multline}
  \label{eq:34}
   \begin{array}{c}
    \langle \varnothing |\\
    \otimes\\
    \langle \varnothing |\\
        \otimes\\
    \langle \varnothing |
  \end{array} \Bigl( \ref{eq:21} \Bigr) \begin{array}{c}
    |\varnothing \rangle\\
    \otimes\\
    |\varnothing \rangle\\
        \otimes\\
    |\varnothing \rangle
  \end{array} \sim \prod_{k<l}^n \frac{\left( \frac{w_k}{w_l} ;q
    \right)_{\infty}}{\left( t \frac{w_k}{w_l} ;q
    \right)_{\infty}}\times\\
  \times  \oint_{\mathcal{C}_{k_1,\ldots, k_n}} d^my
  \oint_{\mathcal{C}_{l_1,\ldots, l_k}} d^kz \prod_{i=1}^m y^{\log_q
    \frac{u_2}{u_1} - 2 \beta - 1} \prod_{i=1}^k z^{\log_q
    \frac{u_3}{u_2} - 2 \beta - 1}\frac{\Delta_k^{(q,t)}(\vec{z}) \Delta_m^{(q,t)}(\vec{y})}{\bar{\Delta}_{k,m}^{(q,t)}(\vec{z},\vec{y})\bar{\Delta}_{m,n}^{(q,t)}(\vec{y},\vec{w})},
\end{multline}
where $\Delta$ and $\bar{\Delta}$ are given in Eq.~\eqref{eq:43}.
The corresponding gauge theory is a linear quiver
\begin{equation}
  \label{eq:44}
     \includegraphics[valign=c]{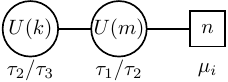}
\end{equation}
For $N$ horizontal lines there will be $(N-1)$ intermediate stacks of
D3 branes and correspondingly, the $3d$ quiver would have $(N-1)$
nodes, since each D3 stack corresponds to a gauge group.

\section{Bosonic and fermionic screenings}
\label{sec:boson-ferm-scre}
In this section we follow the natural development of our formalism and
introduce additional intertwiners, which are obtained by acting with
the $\mathfrak{S}_3$ permutation symmetry of DIM on $\Phi$ and
$\Phi^{*}$. When combined into a network, these new intertwiners
produce new screening currents and charges, defining more general
$W$-algebras and $3d$ quiver gauge theories, which both turn out to be
associated with super algebras. In this way we reproduce the results
of~\cite{BFM} on $W$-algebras associated to DIM algebra in a
simplified and streamlined way.

\subsection{Dual screenings}
\label{sec:dual-screenings}
As shown in sec.~\ref{sec:horiz-fock-repr}, the Fock space
representation $\mathcal{F}_u^{q,t^{-1}}$ of DIM is invariant under
the symmetry $q \leftrightarrow t^{-1}$. However, the intertwining
operators, $\Phi$ and $\Phi^{*}$ from Eqs.~\eqref{eq:7}
and~\eqref{eq:5} respectively are not invariant. This is not a
surprise of course, since the operators $\Phi$ and $\Phi^{*}$
intertwine tensor products of $\mathcal{F}_u^{q,t^{-1}}$ with $V_q^w$
and while the former is invariant under $q \leftrightarrow t^{-1}$,
the latter is not. Acting with the symmetry $q \leftrightarrow t^{-1}$
on $\Phi$ and $\Phi^{*}$ we obtain \emph{new} intertwiners. At this
point we need to refine our notation slightly in order not to confuse
different operators. We call the intertwiners $\Phi$ and $\Phi^{*}$
from Eqs.~\eqref{eq:7} and~\eqref{eq:5} $\Phi^q_{q,t^{-1}}(w)$ and
$\Phi_q^{*q,t^{-1}}(w)$ with the indices signifying the spaces they
act on. We also introduce the color-coded graphical notation with
three vector representations $\mathcal{V}^q_w$,
$\mathcal{V}^{t^{-1}}_w$ and $\mathcal{V}^{t/q}_w$ drawn as blue, red
and violet dashed lines respectively. In this way each parameter
$(q,t^{-1},t/q)$ corresponds to a color: $q$ to blue, $t^{-1}$ to red,
$\frac{t}{q}$ to violet. The Fock spaces also carry the color
corresponding to their ``missing index'': $\mathcal{F}_u^{q,t^{-1}}$
is violet, $\mathcal{F}_u^{q,t/q}$ is red,
$\mathcal{F}_u^{t^{-1},t/q}$ is blue. The intertwiners
$\Phi^q_{q,t^{-1}}(w)$ and $\Phi_q^{*q,t^{-1}}(w)$ are then drawn as
\begin{align}
  \label{eq:49}
  \Phi^q_{q,t^{-1}}(w) &= \includegraphics[valign=c]{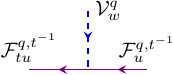}\\
  \Phi_q^{*q,t^{-1}}(w)
  &= \includegraphics[valign=c]{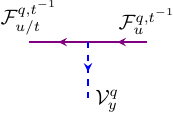} \label{eq:50}
\end{align}
The new intertwiners are drawn as
\begin{align}
  \label{eq:45}
\includegraphics[valign=c]{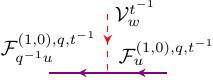}  \quad &=\quad   \Phi^{t^{-1}}_{q,t^{-1}}(w): \mathcal{F}^{(1,0),q,t^{-1}}_u
  \otimes \mathcal{V}^{t^{-1}}_w \to \mathcal{F}_{q^{-1}u}^{(1,0),q,t^{-1}}\\
\includegraphics[valign=c]{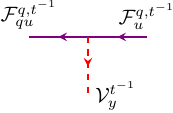}  \quad &=\quad   \Phi_{t^{-1}}^{*q,t^{-1}}(w): \mathcal{F}^{(1,0),q,t^{-1}}_u
 \to \mathcal{F}_{qu}^{(1,0),q,t^{-1}} \otimes \mathcal{V}^{t^{-1}}_w \label{eq:50}
\end{align}
The explicit expressions for the new intertwiners are as follows:
\begin{align}
  \label{eq:46}
  \Phi^{t^{-1}}_{q,t^{-1}}(w) &=e^{- \epsilon_1 Q}
  w^{\frac{P}{\epsilon_2}} \exp \left[ - \sum_{n\geq 1} \frac{w^n}{n}
    a_{-n} \right] \exp \left[ \sum_{n\geq 1} \frac{w^{-n}}{n}
    a_n \right],\\
  \Phi_{t^{-1}}^{*q,t^{-1}}(w) &= e^{\epsilon_1 Q} y^{\frac{1}{\beta}
    - \frac{P}{\epsilon_2}} \exp \left[ \sum_{n\geq 1} \frac{y^n}{n}
    \left( \frac{t}{q} \right)^{\frac{n}{2}} a_{-n} \right] \exp
  \left[ - \sum_{n\geq 1} \frac{y^{-n}}{n} \left( \frac{t}{q}
    \right)^{\frac{n}{2}} a_n \right],
\end{align}
where we have applied the symmetry~\eqref{eq:26} to the
intertwiners~\eqref{eq:7} and~\eqref{eq:5}. Of course, all the results
for normal ordering, commutation and gluing obtained in
sec.~\ref{sec:intertw-ruijs-hamilt} hold for the red dashed lines as
well, provided one exchanges $q \leftrightarrow t^{-1}$ in all the
expressions (hence, for example, the integrations over the spectral
parameters of the intermediate red dashed lines are $t^{-1}$-Jackson
integrals and the irrelevant prefactors are $t$-periodic).

One can use the new intertwiners together with the ``old'' ones to
obtain more general networks. The color rule for gluing the colored
lines is simple: the intertwiners we have just described
(Eqs.~\eqref{eq:49}--\eqref{eq:50}) connect dashed lines and solid
lines of colors which \emph{do not coincide.}  For example, a red
dashed line can connect to blue and violet, but not red horizontal
lines. Let us for the moment consider horizontal lines of only one
color, say violet (the general setup will be described in
sec.~\ref{sec:gluing-differen-fock}).

For the simplest example, consider two violet horizontal lines. We can
stretch either blue or red dashed lines between them:
\begin{multline}
  \label{eq:51}
  \includegraphics[valign=c]{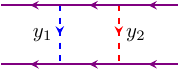} \quad
  = \quad = \int_{-\infty}^{\infty} d_{t^{-1}}y_1
  \int_{-\infty}^{\infty}  d_qy_2\,   \begin{array}{c}
    \Phi^{*q,t^{-1}}_q(y_1)\\
    \otimes\\
    \Phi^{q}_{q,t^{-1}} (y_1)
  \end{array} \begin{array}{c}
    \Phi^{*q,t^{-1}}_{t^{-1}}(y_2)\\
    \otimes\\
    \Phi^{t^{-1}}_{q,t^{-1}} (y_2)
  \end{array} =\\
  =\int_{-\infty}^{\infty} d_{t^{-1}}y_1 \int_{-\infty}^{\infty}  d_qy_2\, S^{q,t^{-1}}_q(y_1) S^{q,t^{-1}}_{t^{-1}}(y_2).
\end{multline}
These two types of lines give rise to two types of screening currents
$S^{q,t^{-1}}_q(w)$ (``blue'' current) and $S^{q,t^{-1}}_{t^{-1}}(w)$
(``red'' current). Both of them commute with the action of DIM
algebra, and therefore with the action of the $q$-Virasoro acting on
the Fock spaces. In fact these two screenings also commute \emph{with
  each other} and constitute the well-known standard set of screenings
of the $q$-Virasoro algebra.

To give a more familiar example of the same situation consider the
ordinary Virasoro algebra built from a free field $\phi(x)$ and
generated by $T(z) = (\partial \phi(z))^2 +
\frac{b-b^{-1}}{2} \partial^2 \phi(z)$. Then there are two stanrard
screening currents $:e^{b \phi(x)}:$ and $:e^{\frac{1}{b} \phi(x)}:$
commuting with $T(z)$ related by the symmetry $b \leftrightarrow
\frac{1}{b}$. This symmetry is exactly the symmetry $q \leftrightarrow
t^{-1}$ of the $q$-deformed model.

As noted in~\cite{BFM}, the red and blue screening currents commute:
\begin{equation}
  \label{eq:52}
  [S^{q,t^{-1}}_q(w), S^{q,t^{-1}}_{t^{-1}}(y)] = 0.
\end{equation}
However, their normal ordering is nontrivial, though it doesn't
contain $q$-Pochhammer symbols as the ordering between screening
currents of the same color:
\begin{equation}
  \label{eq:53}
   S^{q,t^{-1}}_{t^{-1}}(y_1) S^{q,t^{-1}}_q(y_2) = \frac{y_1^2}{\left(
       1 - \frac{1}{t} \frac{y_2}{y_1} \right) \left(
       1 - \frac{1}{q} \frac{y_2}{y_1} \right)} :S^{q,t^{-1}}_{t^{-1}}(y_1) S^{q,t^{-1}}_q(y_2):
\end{equation}
Thus, the integrals corresponding to the networks with two violet
horizontal lines can be described as a pair of coupled
$q$-Dotsenko-Fateev-type integral ensembles:
\begin{multline}
  \label{eq:54}
  \includegraphics[valign=c]{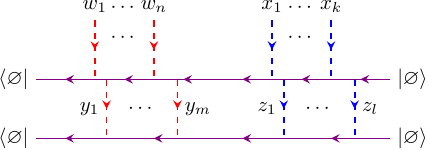}
  \quad \sim\\
  \sim \oint_{\mathcal{C}_{k_1,\ldots, k_n}} d^my
  \oint_{\mathcal{C}_{l_1,\ldots, l_k}} d^l z     \prod_{i=1}^m y^{-\log_t
    \frac{u_2}{u_1} - \frac{2}{\beta} - 1}
  \frac{\Delta_m^{(t^{-1},q^{-1})}(\vec{y})}{\bar{\Delta}_{m,n}^{(t^{-1},q^{-1})}(\vec{y},\vec{w})} \prod_{i=1}^l z^{\log_q
    \frac{u_2}{u_1} - 2 \beta - 1}
  \frac{\Delta_l^{(q,t)}(\vec{z})}{\bar{\Delta}_{l,k}^{(q,t)}(\vec{z},\vec{x})}\\
  \prod_{i=1}^l \prod_{j=1}^m \frac{1}{\left(
       z_i - \frac{1}{t} y_j \right) \left(
       z_i - \frac{1}{q} y_j \right)} \prod_{j=1}^m \prod_{i=1}^k \left( y_j -
     \sqrt{qt} x_i \right) \prod_{j=1}^l \prod_{i=1}^n \left( z_j -
     \sqrt{qt} w_i \right)
\end{multline}
The coupling between two DF integral (one with parameters $(q,t)$ and
the other with $(t^{-1},q^{-1})$) in the last line in
Eq.~\eqref{eq:54} is due to the interaction~\eqref{eq:53}.

From the field theory point of view the integral~\eqref{eq:54}
describes \emph{two} $3d$ theories
\begin{enumerate}
\item $\mathcal{N}=2^{*}_t$ $U(m)$ gauge theory with $n$ fundamental
  multiplets in the $\Omega$-background with parameter $q$ and axial
  mass deformation $t$,

\item $\mathcal{N}=2^{*}_{q^{-1}}$ $U(l)$ gauge theory with $k$
  fundamental multiplets in the $\Omega$-background with parameter
  $t^{-1}$ and axial mass deformation $q^{-1}$.
\end{enumerate}
These two theories are coupled through a $1d$ interaction term. This
setup was described in~\cite{Nieri:2017ntx}.

The situation with more than two violet horizontal lines is similar:
there are twice as many types of screening charges as there were in
sec.~\ref{sec:qw-algebra-scre}. All these screening charges commute
with the action of $qW_N$ algebra. The corresponding field theory is a
pair of $3d$ theories coupled thorough $1d$ interaction.

\subsection{Gluing different Fock spaces. Fermionic screenings}
\label{sec:gluing-differen-fock}
The final logical step in our formalism is to stack together
horizontal lines of different colors. Let us start with two lines,
e.g.\ violet and red. Between them we can stretch dashed lines of the
color different from \emph{both} horizontal lines. There is,
therefore, only \emph{one} choice, blue, which produces the screening
current $S^{q,t^{-1}|q,t/q}_q(w)$
\begin{equation}
  \label{eq:55}
    \includegraphics[valign=c]{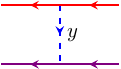}
    \quad =\quad \int_{-\infty}^{\infty} d_qy\, \begin{array}{c}
    \Phi^{*q,t/q}_q(y)\\
    \otimes\\
    \Phi^q_{q,t^{-1}} (y)
  \end{array}   = \int_{-\infty}^{\infty} d_qy\, S^{q,t^{-1}|q,t/q}_q(y).
\end{equation}
Evaluating the screening current explicitly we get
\begin{multline}
  \label{eq:56}
  S^{q,t^{-1}|q,t/q}_q(y) = e^{-(\epsilon_1+\epsilon_2) Q_1
    -\epsilon_2 Q_2} y^{ \frac{\epsilon_1+\epsilon_2- P_1 + P_2}{\epsilon_1}} \exp \left[
    \sum_{n \geq 1} \frac{y^n}{n(1-q^n)} \left( t^{-\frac{n}{2}}
      \left( 1 - (t/q)^n\right) a_{-n}^{(1)} - (1-t^{-n})
      a_{-n}^{(2)}\right)
  \right]\times\\
  \times \exp \left[\sum_{n \geq 1} \frac{y^{-n}}{n(1-q^{-n})} \left(
      - t^{-\frac{n}{2}} \left( 1 - (q/t)^n\right) a_n^{(1)} +
      (1-t^n) a_n^{(2)}\right) \right]
\end{multline}
where the bosons satisfy 
\begin{equation}
  \label{eq:57}
  [a_n^{(1)},a_m^{(1)}] = n \frac{1-q^{|n|}}{1 - \left( \frac{q}{t}
    \right)^{|n|}} \delta_{n+m,0},\qquad   [a_n^{(2)},a_m^{(2)}] = n \frac{1-q^{|n|}}{1 - t^{|n|}} \delta_{n+m,0},
\end{equation}
An explicit calculation shows that the currents
$S^{q,t^{-1}|q,t/q}_q(y)$ \emph{anticommute.} Such fermionic
screenings appear naturally in the context of $W$-algebras associated
with superalgebras~\cite{BFM}. In particular, on two horizontal lines
of different colors, DIM algebra is expected to act as a $W$-algebra
associated to superalgebra $\mathfrak{gl}_{1|1}$ (somewhat similarly
to the case of two Fock spaces of the same color where a product of
$q$-Virasoro and Heisenberg algebras, associated to $\mathfrak{gl}_2$
acts). Having the fermionic screening we can build interesting DF-type
integrals\footnote{Unfortunately the name superintegral is already
  taken.} and the corresponding $3d$ quiver gauge theories. Both of
them are associated to Dynkin diagrams of superalgebras. Let us write
down the simplest example, corresponding to the network
\begin{equation}
  \label{eq:59}
      \includegraphics[valign=c]{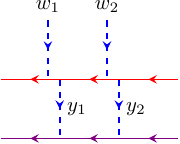}\qquad 
      \Leftrightarrow \qquad       \includegraphics[valign=c]{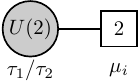}
\end{equation}
The vacuum matrix element of~\eqref{eq:59} is
\begin{equation}
  \label{eq:60}
  \oint_{\mathcal{C}_{1,1}}  (y_1 y_2)^{\log_q
    \frac{u_2}{u_1} + 2 \beta - 3}
  \frac{\Delta_2(\vec{y})}{\bar{\Delta}_{2,2}^{(q,q/t)}(\vec{y},\vec{w})}, 
\end{equation}
where $\Delta_2(\vec{y})$ is the square of the \emph{ordinary} (i.e.\
not $(q,t)$-deformed) Vandermonde determinant:
\begin{equation}
  \label{eq:61}
  \Delta_m(\vec{y}) = \prod_{i\neq j}^m \left( 1- \frac{y_i}{y_j} \right).
\end{equation}
This gives us a hint at what a $3d$ $\mathcal{N}=2^{*}$ theory
associated to superquiver looks like: the gauge node (the only one in
the example, painted graw in~\eqref{eq:59}), corresponding to the
fermionic root has \emph{trivial} axial deformation parameter, as if
$\mathcal{N}=4$ supersymmetry was unbroken. More generally, for
several bosonic roots separated by a fermionic one, the axial mass
parameter \emph{changes sign} along the quiver: e.g.\ it is $t$ for
the bosonic nodes to the left of the fermionic node, and $\frac{q}{t}$
for the bosonic nodes to the right of the fermionic one. On the
fermionic node we effectively have the theory with $t=q$.

Finally, as an exercise we draw a colorful picture incorporating
various screenings we have obtained:
\begin{equation}
  \label{eq:58}
   \includegraphics[valign=c]{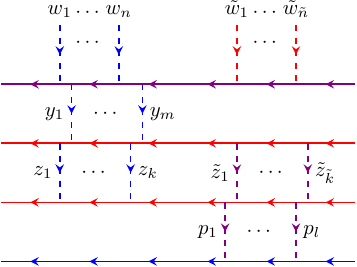} 
\end{equation}
The resulting $qW$-algebra~\cite{BFM} should be associated to a
certain reduction of a sum of superalgebras $\mathfrak{gl}_{1|2}
\oplus \mathfrak{gl}_{2|1}$. The field theory system corresponding to
this brane diagram consists of \emph{three} $3d$ theories each living
in its own $S^1\times \mathbb{R}^2$ space with $\Omega$-background
parameters $q$, $t^{-1}$ and $\frac{t}{q}$ respectively:
\begin{enumerate}
\item Supersymmetric quiver theory associated to the algebra
  $\mathfrak{gl}_{2|1}$ on $S^1\times_q \mathbb{R}^2$ with axial mass
  parameters $t$ and $\frac{q}{t}$.

\item Supersymmetric quiver theory associated to the algebra
  $\mathfrak{gl}_{1|1}$ on $S^1\times_{\frac{t}{q}} \mathbb{R}^2$ with
  axial mass parameters $q^{-1}$ and $t$.

\item Theory of free multiplets on $S^1\times_{t^{-1}} \mathbb{R}^2$
  with axial mass deformation $q^{-1}$.
\end{enumerate}
Pairs of theories $(1,2)$ and $(2,3)$ are coupled by a $1d$
interaction terms.

\section{Ruijsenaars Hamiltonians and their supersymmetric versions}
\label{sec:ruijs-hamilt}
In this section we prove that the networks we have constructed are the
eigenfunctions of (supersymmetric) Ruijsenaars Hamiltonians acting on
the spectral parameters of the external dashed lines.

We first consider a simple example and then argue that the statement
actually holds for a larger class of networks. Our example is the
$T[SU(2)]$ theory. The proof here is already well-known, but we
rederive it using the intertwining property of Higgsed networks to
show that it is automatic in our approach. Consider the action of DIM
element $\Delta^3(x_0^{+})$ on the corresponding
network~\eqref{eq:35}:
\begin{equation}
  \label{eq:62}
   \includegraphics[valign=c]{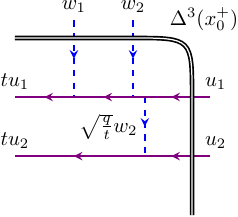}\qquad =\qquad    \includegraphics[valign=c]{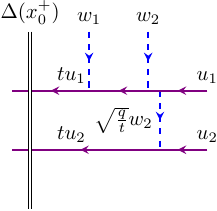} 
 \end{equation}
 where the equality follows from the definition of the intertwining
 operator. Using the coproduct from sec.~\ref{sec:coproduct} we find
 that on the tensor product of two vector representation and two Fock
 representations one has
\begin{multline}
  \label{eq:63}
  \Delta^3(x_0^{+})|_{\mathcal{V}^q_{w_1}\otimes
    \mathcal{V}^q_{w_2}\otimes \mathcal{F}^{q,t^{-1}}_{u_1}\otimes
    \mathcal{F}^{q,t^{-1}}_{u_1}} =
  \Delta(x_0^{+})|_{\mathcal{V}^q_{w_1}\otimes
    \mathcal{V}^q_{w_2}} +\\
  +\oint_{\mathcal{C}} \frac{dz}{z} (\psi^{-}(z) \otimes
  \psi^{-}(z))|_{\mathcal{V}^q_{w_1}\otimes \mathcal{V}^q_{w_2}}
  \otimes \Delta(x^{+}(z))|_{\mathcal{F}^{q,t^{-1}}_{u_1}\otimes
    \mathcal{F}^{q,t^{-1}}_{u_1}}
\end{multline}
where the integration contour $\mathcal{C}$ is a small contour around
$z=0$. Eq.~\eqref{eq:63} gives the DIM action featuring in the l.h.s.\
of Eq.~\eqref{eq:62}.

DIM action in the r.h.s.\ of Eq.~\eqref{eq:62} is just
$\Delta(x^{+}(z))|_{\mathcal{F}^{q,t^{-1}}_{u_1}\otimes
  \mathcal{F}^{q,t^{-1}}_{u_1}}$ (see~\eqref{eq:23}). Sandwiching both
sides of Eq.~\eqref{eq:62} between the vacuum states, we find that
$\Delta(x^{+}(z))|_{\mathcal{F}^{q,t^{-1}}_{u_1}\otimes
  \mathcal{F}^{q,t^{-1}}_{u_1}} |\varnothing\rangle \otimes
|\varnothing\rangle $ contains only \emph{non-negative} powers of $z$,
as does $(\psi^{-}(z) \otimes
\psi^{-}(z))|_{\mathcal{V}^q_{w_1}\otimes \mathcal{V}^q_{w_2}}$ (see
the representations in sec.~\ref{sec:horiz-fock-repr}
and~\ref{sec:vert-vect-repr}). Thus, the integral in Eq.~\eqref{eq:63}
is equal to the zero mode of
$\Delta(x^{+}(z))|_{\mathcal{F}^{q,t^{-1}}_{u_1}\otimes
  \mathcal{F}^{q,t^{-1}}_{u_1}} |\varnothing\rangle \otimes
|\varnothing\rangle $, i.e.\ to $\frac{(u_1 + u_2)}{(1-q^{-1})(1-t)}
|\varnothing\rangle \otimes |\varnothing\rangle$. We have
      \begin{multline}
        \label{eq:3}
        \Delta^3(x_0^{+})|_{\mathcal{V}^q_{w_1}\otimes
          \mathcal{V}^q_{w_2}\otimes
          \mathcal{F}^{q,t^{-1}}_{u_1}\otimes
          \mathcal{F}^{q,t^{-1}}_{u_1}} |w_1\rangle \otimes
        |w_2\rangle \otimes |\varnothing\rangle \otimes
        |\varnothing\rangle = - \frac{1}{1-q^{-1}} |q w_1\rangle \otimes |w_2\rangle
        \otimes |\varnothing\rangle \otimes
        |\varnothing\rangle+\\
        -\frac{1}{1-q^{-1}} \frac{\left( 1 - t \frac{w_2}{w_1} \right)\left( 1 -
            \frac{q}{t} \frac{w_2}{w_1} \right)}{\left( 1 -
            \frac{w_2}{w_1} \right)\left( 1 - q \frac{w_2}{w_1}
          \right)} |w_1\rangle \otimes |q w_2\rangle \otimes
        |\varnothing\rangle \otimes |\varnothing\rangle + \frac{u_1 +
        u_2}{(1-q^{-1})(1-t)} |w_1\rangle \otimes |w_2\rangle \otimes
        |\varnothing\rangle \otimes |\varnothing\rangle
      \end{multline}
      In the r.h.s.\ of Eq.~\eqref{eq:62} we find that $\langle
      \varnothing | \otimes \langle \varnothing | \Delta(x^{+}_0)$
      also reduces to zero modes, which in this case are $\langle
      \varnothing | \otimes \langle \varnothing | \frac{(tu_1
        + tu_2)}{(1-q^{-1})(1-t)}$. Substituting both sides of
      Eq.~\eqref{eq:62} and denoting the vacuum matrix element of the
      network by $\psi(\vec{w}, \vec{u})$ we can write
      \begin{equation}
        \label{eq:47}
        \left( q^{w_1 \partial_{w_1}} +  \frac{\left( 1 - t \frac{w_2}{w_1} \right)\left( 1 -
            \frac{q}{t} \frac{w_2}{w_1} \right)}{\left( 1 -
            \frac{w_2}{w_1} \right)\left( 1 - q \frac{w_2}{w_1}
          \right)} q^{w_2 \partial_{w_2}} \right)\psi(\vec{w}, \vec{u}) = (u_1
        + u_2) \psi(\vec{w}, \vec{u})
      \end{equation}
      Factoring the flipping field contributions (see
      Eq.~\eqref{eq:33}) out of $\psi(\vec{w}, \vec{u})$ we find that
      the function
      \begin{equation}
        \label{eq:48}
        \tilde{\psi}(\vec{w},
      \vec{u}) = \frac{\left( t
          \frac{w_1}{w_2} ;q \right)_{\infty}}{\left( 
          \frac{w_1}{w_2} ;q \right)_{\infty}} \psi(\vec{w},
      \vec{u})
    \end{equation}
    is an eigenfunction of the Ruijsenaars Hamiltonian:
    \begin{equation}
      \label{eq:64}
      H_1\tilde{\psi}(\vec{w},
      \vec{u}) = \left( \frac{1- t \frac{w_1}{w_2}}{1 - \frac{w_1}{w_2}} q^{w_1 \partial_{w_1}} +  \frac{ 1 - t \frac{w_2}{w_1} }{ 1 -
            \frac{w_2}{w_1}} q^{w_2 \partial_{w_2}} \right)\tilde{\psi}(\vec{w}, \vec{u}) = (u_1
        + u_2) \tilde{\psi}(\vec{w}, \vec{u}).
    \end{equation}
The argument of intertwining is completely general, thus any Higgsed
network of the form we have considered is an eigenfunction of the
Ruijsenaars Hamiltonians with eigenvalues determined by the spectral
parameters of the horizontal lines.

Let us finally mention the supersymmetric generalization of the
Ruijsenaars system, which is obtained by considering the
network~\eqref{eq:54}. In this case there are two types of DIM action
in the vertical representations, which give rise to two coupled
Ruijsenaars Hamiltonians:
\begin{equation}
  \label{eq:65}
  H_1^{\mathrm{super}} = \sum_{i=1}^k \prod_{a=1}^n \frac{x_i - q w_a}{x_i - w_a}
  \prod_{j\neq i} \frac{t x_i- x_j}{x_i- x_j} q^{x_i \partial_{x_i}} + \frac{1-q^{-1}}{1-t}
  \sum_{i=1}^n \prod_{a=1}^k \frac{w_i - t^{-1} x_a}{w_i - x_a}
  \prod_{j\neq i} \frac{q^{-1} w_i- w_j}{w_i- w_j} t^{-w_i \partial_{w_i}}
\end{equation}
The interaction terms in the Hamiltonian are due to the nontriviality
of the DIM coproduct. Denoting the value of the vacuum matrix element
of~\eqref{eq:54} (stripped off the flipping fields) by
$\tilde{\psi}(\vec{w},\vec{x},\vec{u})$ we get
\begin{equation}
  \label{eq:66}
  H_1^{\mathrm{super}} \tilde{\psi}(\vec{w},\vec{x},\vec{u}) =
  \frac{1}{1-t}\left( (1 - t^{k-l}q^{m-n}) u_1 + (1 - t^lq^m) u_2\right)\tilde{\psi}(\vec{w},\vec{x},\vec{u}).
\end{equation}

Such system was considered in~\cite{FS}\footnote{Their notation
  differs from the notation here: $(q,t)_{\cite{FS}} = (q,t^{-1})_{\mathrm{here}}$}. Our formalism provides
explicit integral formulas for the solutions of the supersymmetric
Ruijsenaars system and of the corresponding supersymmetric Macdonald
polynomials (and, in the $t \to q$ limit, supersymmetric Schur
functions).

\section{Conclusions and discussions}
\label{sec:concl-disc}
We have introduced a version of the network formalism based on the DIM
algebra intertwiners. It can be thought of as an analogue of the
refined topological vertex formalism for the case incorporating not
only a network of five-branes of Type IIB, but also D3 branes. On the
field theory side it provides a way to understand partition functions
(holomorphic blocks) of $3d$ $\mathcal{N}=2^{*}$ quiver gauge
theories. In particular, it gives a constructive definition of the
partition functions of the theories associated with Dynkin diagrams
for superalgebras and gives a transparent proof of the fact that the
partition functions are eigenfunctions of (supersymmetric)
Ruijsenaars-Schneider Hamiltonians. On the algebraic side our
construction naturally produces screenings for the $qW$-algebras,
including those associated with superalgebras.

There are many directions along which one can extend the approach
presented here. For example we have just started studying the $3d$
theories corresponding to superquivers --- there is a wealth of
interesting examples which can be built rather straightforwardly with
a simple set of building blocks, as e.g.\ the picture~\eqref{eq:58}
demonstrates. Some of the networks we have introduced can be
compactified, i.e.\ drawn on a cylinder or torus instead of a
plane. This should give rise to explicit description of the
eigenfunctions of quantum elliptic and double elliptic
systems~\cite{Aminov:2014wra}. It would be interesting to study the
monodromy problems for the networks we have introduce --- they should
be described by the elliptic stable envelopes
theory~\cite{Aganagic:2016jmx}.

\section*{Acknowledgements}
\label{sec:acknowledgements}
The author would like to thank B.~Feigin for a discussion. The author
is supported by the RSF grant 18-71-10073.

\appendix
\section{DIM algebra and its representations}
\label{sec:dim-algebra-its}
For the sake of completeness in this Appendix we list the relevant
formulas from the theory of DIM algebras, mostly taken
from~\cite{AFS} \cite{CP1}.

\subsection{The algebra}
\label{sec:algebra}
DIM algebra $U_{q,t}(\widehat{\widehat{\mathfrak{gl}}}_1)$ is
generated by the currents $x^{\pm}(z) = \sum_{n \in \mathbb{Z}}
x^{\pm}_n z^{-n}$, $\psi^{\pm}(z) = \sum_{n \gtreqless 0} \psi^{\pm}_n
z^{-n}$ and a central element $\gamma$ subject to the
relations\footnote{This definition differs from the definition
  of~\cite{AFS} by the rescaling of the generators
  $x^{\pm}_{\mathrm{our}}(z) = (1-q^{\mp 1})^{-1}(1-t^{\pm 1})^{-1}
  x^{\pm}_{\mathrm{AFS}}(z)$, while keeping
  $\psi^{\pm}_{\mathrm{our}}(z) = \psi^{\pm}_{\mathrm{AFS}}(z)$.}
\begin{gather}
  \label{eq:22}
  [\psi^{\pm}(z), \psi^{\pm}(w)] = 0,\qquad \psi^{+}(z) \psi^{-}(w) =
  \frac{g\left(\gamma \frac{w}{z}\right)}{g\left(\gamma^{-1}
      \frac{w}{z}\right)}
  \psi^{-}(w) \psi^{+}(z),\\
  \psi^{+}(z) x^{\pm}(w) = g\left(\gamma^{\mp \frac{1}{2}}
    \frac{w}{z}\right)^{\mp 1} x^{\pm}(w) \psi^{+}(z), \qquad
  \psi^{-}(z) x^{\pm}(w) = g\left(\gamma^{\mp \frac{1}{2}}
    \frac{w}{z}\right)^{\pm 1} x^{\pm}(w) \psi^{-}(z),\\
  [x^{+}(z), x^{-}(w)] = \frac{1}{G^{-}(1)} \left( \delta \left(
      \gamma^{-1} \frac{z}{w} \right) \psi^{+} \left(
      \gamma^{\frac{1}{2}} w \right) - \delta \left( \gamma
      \frac{z}{w} \right) \psi^{-} \left(
      \gamma^{-\frac{1}{2}} w \right) \right),\\
  G^{\mp}\left( \frac{z}{w} \right) x^{\pm}(z) x^{\pm}(w) =
  G^{\pm}\left( \frac{z}{w} \right) x^{\pm}(w) x^{\pm}(z),\label{eq:24}
\end{gather}
where $\delta(x) = \sum_{n \in \mathbb{Z}} x^n$ and the ``structure
functions''\footnote{Curiously, they also appear as factorized
  scattering matrices of some $2d$ integrable models. It is tempting
  to try to understand DIM algebra as a version of the
  Zamolodchikov-Fateev algebra of such models.} of the algebra are
given by $G^{\pm}(x) = (1 - q^{\pm 1}x)(1 - t^{\mp 1}x)(1 - t^{\pm 1}
q^{\mp 1}x)$ and $g(x) = \frac{G^{+}(x)}{G^{-}(x)}$. Notice that $g
\left( \frac{1}{x} \right) = \frac{1}{g(x)}$ and in particular
$G^{+}(1) = G^{-}(1)$. The ratio of the zero modes
$\frac{\psi^{-}_0}{\psi^{+}_0} = \gamma_{\perp}^2$ also turns out to
be central\footnote{The \emph{product} of the zero modes $\psi^{-}_0
  \psi^{+}_0$ is central too, but can be eliminated by an overall
  rescaling of \emph{all} the currents.}. There are also Serre
relations for triple commutators of $x^{+}(z)$ and $x^{-}(z)$, which
we will not write down here.

Notice that the relations of the algebra are \emph{manifestly}
symmetric under the action of $\mathfrak{S}_3$ group permuting the
triplet of deformation parameters $\left( q, t^{-1}, t/q
\right)$. Only part of this symmetry will be retained by the
representations which we are going to consider. By this we mean that
some permutations will not affect a representation, while others will
turn a representation into an isomorphic one.

DIM algebra respects two gradings, $d$ and $d^{\perp}$. $d$ counts the
number of the Laurent mode of a current, so that $d(x_n^{\pm}) =
d(\psi_n^{\pm}) = n$, while $d^{\perp}$ is a ``perpendicular'' grading
defined as $d^{\perp}(x^{\pm}_n) = \pm 1$, $d^{\perp}(\psi^{\pm}_n) =
0$.

There is an extra symmetry of the DIM algebra, which is not manifest
in the definition~\eqref{eq:22}--\eqref{eq:24} --- the
$SL(2,\mathbb{Z})$ automorphism group (the most tricky part of it is
the action of the $S$-element, known as the Miki automorphism). We
will not define the action of this symmetry on the currents
explicitly. It will be enough for us to visualize it as an action of
$SL(2,\mathbb{Z})$ on the double grading lattice $(d, d^{\perp}) \in
\mathbb{Z}^2$ and on the doublet of central charges $(\gamma^2,
\gamma_{\perp}^2)$, so that e.g.\ $x_0^{+}$ is turned into
$\psi^{-}_{-1}$ by the $S$ element.

\subsection{The coproduct}
\label{sec:coproduct}
DIM algebra can be endowed with a coproduct $\Delta$, which acts on
the generators as follows:
\begin{align}
  \label{eq:23}
  \Delta(x^{+}(z)) &= x^{+}(z) \otimes 1 + \psi^{-}\left(
    \gamma_{(1)}^{\frac{1}{2}}z \right) \otimes x^{+} \left(
    \gamma_{(1)} z \right),\\
  \Delta(x^{-}(z)) &= x^{-}\left(\gamma_{(2)} z\right) \otimes
  \psi^{+}\left( \gamma_{(2)}^{\frac{1}{2}}z \right) + 1 \otimes
  x^{-}(z),\\
  \Delta(\psi^{\pm}(z)) &= \psi^{\pm}\left(
    \gamma_{(2)}^{\pm \frac{1}{2}}z \right) \otimes \psi^{\pm}\left(
    \gamma_{(1)}^{\mp \frac{1}{2}}z \right).
\end{align}
where $\gamma_{(1)}$ (resp.\ $\gamma_{(2)}$) denotes the central
charge of the first (resp.\ second) representation in the tensor
product. Since the currents are formal infinite Laurent series, the
products in the r.h.s.\ may require regularization for some
representations. We will not encounter this problem for the
representations we will consider.

The coproduct $\Delta$ respects the $\mathfrak{S}_3$ permutation
symmetry of the DIM algebra but is \emph{not} invariant under the
action of $SL(2,\mathbb{Z})$. In fact there is an infinite number of
coproducts, parametrized by irrational slopes on the $2d$ plane. All
these coproducts are related to each other by nontrivial Drinfeld
twists (see~\cite{Awata:2016mxc}, and for a more geometric view
also~\cite{Maulik:2012wi}).

\subsection{Horizontal Fock representation}
\label{sec:horiz-fock-repr}
There is a representation of DIM algebra on the Fock space
$\mathcal{F}_u^{(1,0),q,t^{-1}}$, generated by the action of creation
operators $a_{-n}$, $n \in \mathbb{Z}_{> 0}$ on the vacuum vector
$|\varnothing, u\rangle$. The states of $\mathcal{F}_u^{q,t^{-1}}$ are
therefore labelled by Young diagrams. The combination of indices
$(1,0)$ and $q,t^{-1}$ of the representation is used to denote the
``direction'' (horizontal) and ``length'' of its central charge vector
$(\gamma^2, \gamma_{\perp}^2) = ((t/q)^1, (t/q)^0) = (t/q, 1)$
respectively. We will usually omit the index $(1,0)$, when it is clear
what is the \emph{direction} of the central charge vector. Creation
and annihilation operators satisfy the commutation relations
\begin{equation}
  \label{eq:8}
  [a_n, a_m] = n \frac{1 - q^{|n|}}{1-t^{|n|}} \delta_{n+m,0},
\end{equation}
while the zero modes $P$ and $Q$ commute with $a_n$ and satisfy the
standard Heisenberg commutation relations
\begin{equation}
  \label{eq:13}
  [P,Q] = 1.
\end{equation}
The zero modes act on the vacuum vector as follows:
\begin{equation}
  \label{eq:19}
  P |\varnothing, u\rangle = \ln u\, |\varnothing, u\rangle,\qquad
  e^{\alpha Q} |\varnothing,
  u\rangle = |\varnothing, e^{\alpha} u\rangle.
\end{equation}
The action of the DIM generators is given by the following vertex
operators:
\begin{align}
  \label{eq:25}
  x^{+}(z) &= (1-q^{-1})^{-1} (1-t)^{-1} e^P \exp \left[ \sum_{n \geq
      1} \frac{z^n}{n} (1-t^{-n}) a_{-n} \right] \exp \left[ - \sum_{n
      \geq 1} \frac{z^{-n}}{n} (1-t^n) a_n \right],\\
  x^{-}(z) &= (1-q)^{-1} (1-t^{-1})^{-1} e^{-P} \exp \left[ - \sum_{n
      \geq 1} \frac{z^n}{n} (1-t^{-n}) \left( \frac{t}{q}
    \right)^{\frac{n}{2}} a_{-n} \right] \exp \left[ - \sum_{n \geq 1}
    \frac{z^{-n}}{n} (1-t^n) \left( \frac{t}{q}
    \right)^{\frac{n}{2}} a_n \right],\notag\\
  \psi^{+}(z) &= \exp \left[ - \sum_{n\geq 1} \frac{z^{-n}}{n}
    (1-t^n) \left( 1 - \left( \frac{t}{q} \right)^n \right) \left(
      \frac{q}{t} \right)^{\frac{n}{4}} a_n
  \right],\\
  \psi^{-}(z) &= \exp \left[ \sum_{n\geq 1} \frac{z^n}{n}
    (1-t^{-n}) \left( 1 - \left( \frac{t}{q} \right)^n \right) \left(
      \frac{q}{t} \right)^{\frac{n}{4}}  a_{-n}
  \right].\label{eq:27}
\end{align}

The representation $\mathcal{F}_u^{q,t^{-1}}$ is invariant under the
exchange of $q$ and $t^{-1}$ deformation parameters. To see this we
notice that the exchange $q \leftrightarrow t^{-1}$ in the vertex
operators~\eqref{eq:25}--\eqref{eq:27} is equivalent to the rescaling
of the creation and annihilation operators:
\begin{equation}
  \label{eq:26}
  a_n^{(q,t^{-1})} = \frac{1-q^{-n}}{1-t^n}
  a_n^{(t^{-1},q)}.
\end{equation}
Notice in particular, that $a_n^{(t^{-1},q)}$ satisfy
\begin{equation}
  \label{eq:28}
  [a_n^{(t^{-1},q)}, a_m^{(t^{-1},q)}] = n \frac{1 -
    t^{-|n|}}{1-q^{-|n|}} \delta_{m+n,0},
\end{equation}
as they should. This symmetry might be familiar from the theory of
Macdonald polynomials $M^{(q,t^{-1})}_Y(a_{-n}^{(q,t^{-1})})$, in
which it corresponds to the transposition of the Young diagram $Y$. As
we have mentioned above, this $\mathbb{Z}_2$ symmetry of the Fock
representation is part of the larger $\mathfrak{S}_3$ permutation
symmetry of the DIM algebra. The remaining elements of
$\mathfrak{S}_3$ transform $\mathcal{F}_u^{q,t^{-1}}$ into two more
horizontal Fock representations $\mathcal{F}_u^{q,t/q}$ and
$\mathcal{F}_u^{t^{-1}, t/q}$, obtained
from~\eqref{eq:25}--\eqref{eq:27} by permuting the parameters of the
algebra. Their central charge vectors $(\gamma^2, \gamma_{\perp}^2)$
are $(t^{-1}, 1)$ and $(q, 1)$ respectively.

Fock representation with different \emph{slope,} e.g.\ a
\emph{vertical} one $\mathcal{F}_u^{(0,1),q,t^{-1}}$ can be obtained
by the action of the elements of $SL(2,\mathbb{Z})$ automorphism group
on $\mathcal{F}_u^{(1,0),q,t^{-1}}$. We will not need these
representations for the construction presented in the main text, so we
omit their explicit definition. For more information on the action of
$SL(2,\mathbb{Z})$ and its implications
see~\cite{Zenkevich:2014lca},~\cite{Mironov:2016yue},~\cite{Bourgine:2016vsq},~\cite{Morozov:2015xya}.

\subsection{Vertical vector representation}
\label{sec:vert-vect-repr}
The vertical vector representation $\mathcal{V}_w^q$ has trivial
central charges $(\gamma^2, \gamma_{\perp}^2) = (1,1)$. It can be
understood as a kind of \emph{evaluation} representation for the
currents $x^{\pm}(z)$, $\psi^{\pm}(z)$, similar to evaluation
representations of (quantum) affine algebra. There are two equivalent
ways to view $\mathcal{V}_w^q$: either as an infinite dimensional
representation parametrized by $w$ with basis $|n,w\rangle$, $n \in
\mathbb{Z}$, or as a representation on the space of functions of $w$
with state $|n, w\rangle$ corresponding to function $|q^n
w\rangle$. In the first case the action of the DIM currents is
\begin{align}
  x^{+}(z)|w,n\rangle &= -\frac{1}{1-q^{-1}} \delta \left(
    \frac{q^n w}{z} \right)
  |w,n+ 1\rangle, \notag\\
  x^{-}(z)|w,n\rangle &= \frac{1}{1-q} \delta \left(
    \frac{q^{n-1} w}{z} \right)
  |w,n- 1\rangle, \notag\\
  \psi^{+}(z) |w,n\rangle &= \frac{\left( 1 - \frac{t}{q} \frac{q^n
        w}{z}\right) \left( 1 - \frac{1}{t} \frac{q^n
        w}{z}\right)}{\left( 1 - \frac{q^n w}{z} \right) \left( 1 -
      \frac{1}{q} \frac{q^n w}{z} \right)}
  |w,n\rangle, \label{eq:4}\\
  \psi^{-}(z) |w,n\rangle &= \frac{\left( 1 - \frac{q}{t} \frac{z}{q^n
        w}\right) \left( 1 - t \frac{z}{q^n w}\right)}{\left( 1 -
      \frac{z}{q^n w} \right) \left( 1 - q \frac{z}{q^n w} \right)}
  |w,n\rangle. \notag
\end{align}
In the second view the currents act on functions $|w\rangle$:
\begin{align}
  x^{+}(z)|w\rangle &= - \frac{1}{1-q^{-1}} \delta \left( \frac{w}{z} \right)
  |q w \rangle, \notag\\
  x^{-}(z)|w\rangle &= - \frac{1}{1-q} \delta \left( \frac{w}{q z} \right)
  \left| \frac{w}{q} \right\rangle, \notag\\
  \psi^{+}(z) |w\rangle &= \frac{\left( 1 - \frac{t}{q} \frac{
        w}{z}\right) \left( 1 - \frac{1}{t} \frac{w}{z}\right)}{\left(
      1 - \frac{w}{z} \right) \left( 1 - \frac{1}{q} \frac{w}{z}
    \right)}
  |w\rangle,\label{eq:6}\\
  \psi^{-}(z) |w\rangle &= \frac{\left( 1 - \frac{q}{t}
      \frac{z}{w}\right) \left( 1 - t \frac{z}{w}\right)}{\left( 1 -
      \frac{z}{w} \right) \left( 1 - q \frac{z}{w} \right)} |w\rangle.
  \notag
\end{align}
We will use both views interchangeably at our convenience.

The vector representation $V^q_w$ is manifestly symmetric with respect
to the exchange of $t$ and $\frac{q}{t}$. The action of
$\mathfrak{S}_3$ permutation symmetry of DIM algebra produces two more
vector representations $\mathcal{V}^{t^{-1}}_w$ and
$\mathcal{V}^{t/q}_w$ defined in an obvious way.

\subsection{Visualizing representations}
\label{sec:macm-repr-visu}
Let us also mention that both Fock and vector representations can be
thought of as certain reductions of a more general MacMahon
representation of central charge $(1,K)$
\cite{FJMM-plane},~\cite{Zenkevich:2017tnb},~\cite{Awata:2018svb} with
general $K \in \mathbb{C}$ and states labelled by \emph{plane}
partitions, i.e.\ $3d$ Young diagrams (here we do not pay attention to
the \emph{direction} of the central charge vector, focusing only on
its ``magnitude''). As a mnemonic aid, one can view the $3d$
partitions constituting the MacMahon representation as living in a
$\mathbb{Z}^3_{>0}$ space with three coordinate axes associated with
three parameters $(q, t^{-1}, t/q)$ of the DIM algebra. A reduction of
the representation corresponds to restriction to a subset of plane
partitions of specific form:
\begin{enumerate}
\item Fock representation $\mathcal{F}^{q,t^{-1}}_u$ contains plane
  partitions of unit thickness lying along the $(q,t^{-1})$ plane
  inside $\mathbb{Z}_{>0}^3$, i.e.\ those that reduce to Young
  diagrams. It is evident geometrically, that
  $\mathcal{F}^{q,t^{-1}}_u$ is invariant under the exchange of $q$
  and $t^{-1}$ axes and that this symmetry corresponds to the
  transposition of a Young diagram $Y$ labelling a state of the
  representation. There are three coordinate planes, and therefore
  three Fock representations.

\item Vector representation $\mathcal{V}_w^q$ can be visualized as
  single column diagrams towering in the direction associated to
  $q$. One needs to stretch one's imagination a little bit in this
  case, since columns of \emph{negative} height are also
  allowed. Naturally, the representation is invariant with respect to
  the exchange of coordinate axes $t \leftrightarrow \frac{q}{t}$,
  which lie perpendicular to $q$. There are three species of vector
  representations, corresponding to three different orientations of
  the columns.
\end{enumerate}

\end{document}